\documentclass[superscriptaddress,showkeys,showpacs]{revtex4}
\usepackage{graphicx}
\usepackage[tbtags]{amsmath}

\begin{document}
\title{Sea quark and gluon polarization in the nucleon at NLO accuracy.}
\thanks{Partially supported by CONICET, Fundaci\'on Antorchas, UBACYT and 
ANPCyT, Argentina.}
\author{D. de Florian} 
\email{deflo@df.uba.ar}
\affiliation{Departamento de F\'{\i}sica,
Universidad de Buenos Aires\\ Ciudad Universitaria, Pab.1 (1428)
Buenos Aires, Argentina}
\author{G. A. Navarro} 
\email{gabin@df.uba.ar}
\affiliation{Departamento de F\'{\i}sica,
Universidad de Buenos Aires\\ Ciudad Universitaria, Pab.1 (1428)
Buenos Aires, Argentina}
\author{R. Sassot}
\email{sassot@df.uba.ar}
\affiliation{Departamento de F\'{\i}sica,
Universidad de Buenos Aires\\ Ciudad Universitaria, Pab.1 (1428)
Buenos Aires, Argentina}
\date{{\bf \today}}

\begin{abstract}
We investigate the sea quark polarization in the nucleon by means 
of a combined next to leading order analysis to the recently enlarged set of 
inclusive and semi-inclusive polarized deep inelastic scattering data. Using 
the Lagrange multiplier method, we asses the uncertainty inherent to the 
extraction of the different spin dependent parton densities in a QCD global 
fit, and the impact of the increased set of semi-inclusive data now available.
We comment on future prospects at RHIC and JLAB and their potential impact
in the future determination of polarized parton densities.
\end{abstract}

\pacs{12.38.Bx, 13.85.Ni}
\keywords{Semi-Inclusive DIS; perturbative QCD.}

\maketitle

\section{Introduction}
For more than fifteen years, polarized inclusive deep inelastic scattering 
(pDIS) has been the main, if not unique, source of information on how the 
individual partons in the nucleon are polarized at very short distances. 
Hadronic decays, of course, bring us from lower energies hints about the 
relations between the net polarization of each flavor, but only after 
making  strong assumptions about flavor symmetry within the nucleon. These 
kind of assumptions, however, have been seriously challenged by data over the 
last decade, at least in the realm unpolarized experiments \cite{MRST04,CTEQ}. 

The electromagnetic nature of the dominant interaction between the leptonic 
probe and the target in the pDIS experiments performed so far, does not allow 
to disentangle quark from anti-quark contributions, and thus valence quark
from sea quarks. Consequently, in spite of strenuous efforts and  very 
successful experimental programs run by several collaborations, our knowledge 
of the detailed spin dependent parton distribution functions (pPDF) has been 
held hostage to a large extent by our own assumptions. Most notably, the sea 
quark densities and 
also that for the gluons. The latter, because it influences the pDIS observables
mainly through their energy scale dependence. Although at next to leading 
order accuracy the gluon density contributes to the pDIS cross section  
directly, this contribution is relatively suppressed.

Many alternative experiments have been conceived, and in some cases already 
implemented, to improve this situation. The most mature among them are those
based on polarized semi-inclusive deep inelastic scattering (pSIDIS), i.e. a 
pDIS experiment where a particular hadron is tagged in the hadronic final 
state. Choosing different target and final state hadrons, the respective cross 
sections are sensitive to different combinations of flavored quarks and 
anti-quarks, to be disentangled. These experiments began with the pioneering 
efforts of SMC, in the late nineties \cite{SMC}, followed by those of HERMES 
at DESY \cite{HERMES}, and lately COMPASS at CERN \cite{COMPASS}, and are 
planned to be improved at the Thomas Jefferson National Laboratory (JLAB) in 
the near future \cite{xiaodong}.

The phenomenological impact of the pSIDIS data proved to be encouraging 
although scarce in the initial stage: the reduced number of data and the
relatively large estimated errors, at best allowed to check the consistency 
between pDIS and pSIDIS in a variety of spin-flavor symmetry scenarios 
\cite{DSS,DS,NS}. With the 
availability of larger sets of pSIDIS data, much more precise, and for final 
state hadrons and targets of different flavor composition, the situation now 
has changed dramatically. pSIDIS data have a non negligible weight in 
combined global fits at present, comparable to that of inclusive data, and 
also show clear preferences for the light sea quark polarization. It also 
helps to constrain the strange sea quark and gluon polarization complementing 
the information already obtained from pDIS.       

Although we can now leave behind the spin-flavor symmetry assumption, 
translating polarized observables to parton distributions always imply 
the use of some additional piece of information. In the case of pDIS, we need 
the unpolarized structure function $F_1$, or a set of unpolarized parton 
densities (PDFs), in order to go from the pPDFs to the measured asymmetries. 
For pSIDIS we also need a set of fragmentation functions (FFs), the functions 
that measure the probability of a quark fragmenting into a given hadron 
\cite{KKP,kretzer}, and polarized fracture functions \cite{deFlorian:1995fd}
if target fragmentation events are taken into account, or a Monte Carlo 
generator modeling these same processes. 
The results of the global analyses to polarized data show negligible 
differences when using different sets of modern PDFs \cite{MRST02,CTEQ}, 
but the differences are sizable when using different FFs, a feature to be 
taken into account when assessing the uncertainties of the pPDFs. 
Eventually, pSIDIS data will help to constrain FFs, including it along with 
electron-positron annihilation data in the global fits where they come from.

In the following we perform a combined next to leading order analysis to the 
recently updated set of pDIS and pSIDIS data, and present the resulting pPDFs. 
Specifically, we focus on the extraction of sea quark and gluon densities,
analyzing the constraining power of either set of data on the individual
densities. As result, we find not only a complete agreement between pDIS 
and pSIDIS data, but a very useful complementarity, leading to rather well 
constrained densities.   

Using the Lagrange multiplier approach \cite{Stump:2001gu}, we explore the 
profile of the $\chi ^2$
function against different degrees of polarization in each parton flavor. In 
this way we obtain estimates for the uncertainty in the net polarization of 
each flavor, and in the parameters of the pPDFs.  We compare results obtained 
with the two most recent sets of fragmentation functions, and also within the 
leading (LO) and next to leading order (NLO) approximation. The differences 
found using alternative sets of FF are found to be within conservative 
estimates for the uncertainties. Nevertheless, there is a clear preference for 
a given set of FF over the other, shown in a difference of several units in 
the $\chi ^2$ of the respective global fits. In NLO global fits the overall 
agreement between theory and the full set of data is sensibly higher than in 
LO case.  

Finally, we analyze the behavior of the cross section for longitudinally 
polarized proton-proton collisions into neutral pions with a wide range of 
pPDFs sets coming from a rather conservative uncertainty interval. 
This observable is found to be crucially sensitive to the polarized gluon 
density and therefore an invaluable tool. We compute the 
required precision to be reached in the programed experiments in order to 
constrain even further this distribution and also future sets of pPDFs. 
A similar analysis is made for forthcoming pSIDIS data to be obtained at JLAB.


\section{Conventions and data sets}

Throughout the present analysis, we follow the same conventions and definitions
for the polarized inclusive asymmetries and parton densities adopted in 
references \cite{NS,DS,DSS}, however we use more recent inputs, such as 
unpolarized parton densities \cite{MRST02} and the respective values for 
$\alpha_s$. 
In the totally inclusive case, the spin dependent asymmetries are given by \cite{review}
\begin{eqnarray}
A_1^N(x,Q^2)=
\frac{g_1^N(x,Q^2)}{F_1^N(x,Q^2)}\,=\,\frac{g_1^N(x,Q^2)}{F_2^N(x,Q^2)/{ 2x[ 1+R^N(x,Q^2)] } },
\end{eqnarray}
where the inclusive spin-dependent nucleon structure function $g_1^N(x,Q^2)$ can be written at NLO as a convolution between polarized parton densities for quarks and gluons, $\Delta q_i(x,Q^2)$ and $\Delta g(x,Q^2)$, respectively, and coefficient functions $\Delta C_i(x)$\cite{inclusiva}
\begin{eqnarray}
g_1^N(x,Q^2)=\frac{1}{2}\sum_{q,\bar{q}}e_q^2 \, 
\Bigg[ \Delta q(x,Q^2)\, +\,\frac{\alpha_s(Q^2)}{2\pi} \int_x^1 \frac{dz}{z} \Bigg\{\Delta C_q(z) \Delta q(\frac{x}{z},Q^2)
+\, \Delta C_g(z) \Delta g(\frac{x}{z},Q^2)\Bigg\}\Bigg].
\end{eqnarray}

A more detailed discussion about these coefficient functions and their factorization scheme dependence can be found in Ref.\cite{newgr}. $F_1^N(x,Q^2)$ is the unpolarized nucleon structure function that can be written in terms of $F_2^N(x,Q^2)$ and R, the ratio of the longitudinal to transverse cross section 
\cite{review}. The use of the QCD NLO approximations both in $F_1^N(x,Q^2)$
and in $g_1^N(x,Q^2)$ has  been shown to reduce the sensitivity of PDFs to 
higher twists \cite{Leader:2005kw}.

Analogously, for the semi-inclusive asymmetries we have:
\begin{eqnarray}
A_1^{Nh}(x,Q^2)\mid_Z\, \simeq \,\frac{\int_Z \,dz\, g_1^{Nh}(x,z,Q^2)}{\int_Z \,dz\, F_1^{Nh}(x,z,Q^2)},
\end{eqnarray}
where the superscript $h$ denotes the hadron detected in the final state, and the variable $z$ is given by the ratio between the hadron energy and that of the spectators in the target. The region $Z$, over which $z$ is integrated, is determined by kinematical cuts applied when measuring the asymmetries.

For the spin dependent structure 
function $g^N_1(x,Q^2)$, we use the NLO expression \cite{NPB}
\begin{eqnarray}
 g_{1}^{N\,h}(x,z,Q) 
%
&=& 
\frac{1}{2}\sum_{q,\bar{q}} \!\!\!\!\!\ e_q^2 \;
\Bigg[ \Delta q \left(x,Q^2\right) D_q^H\left(z,Q^2\right) +
\frac{\alpha_s(Q^2)}{2\pi}
\int_x^1 \frac{d\hat{x}}{\hat{x}} \int_z^1 \frac{d\hat{z}}{\hat{z}}
\Bigg\{
\Delta q \left(\frac{x}{\hat{x}},Q^2\right)
\Delta C_{qq}^{(1)}(\hat{x},\hat{z},Q^2)
D_{q}^H\left(\frac{z}{\hat{z}},Q^2\right) 
\nonumber \\
&&+\Delta q \left(\frac{x}{\hat{x}},Q^2\right)
\Delta C_{gq}^{(1)} (\hat{x},\hat{z},Q^2)
D_{g}^H\left(\frac{z}{\hat{z}},Q^2\right) +
\Delta {g} \left(\frac{x}{\hat{x}},Q^2\right)
\Delta C_{qg}^{(1)} (\hat{x},\hat{z},Q^2)
D_{q}^H\left(\frac{z}{\hat{z}},Q^2\right)\Bigg\}\Bigg],
\end{eqnarray}
and in order to avoid the convolution integrals in $\hat{x}$ and $\hat{z}$
we switch to moment space in both variables as suggested in \cite{SW} and 
already implemented in \cite{NS}.

For $u$ and $d$ quarks {\em plus} anti-quarks densities at the 
initial scale $Q^2_0 =0.5$ GeV$^2$ we propose
\begin{eqnarray}
 x (\Delta q+\Delta \overline{q})  =  N_{q}
\frac{x^{\alpha_q}(1-x)^{\beta_q}(1+\gamma_q\,
x^{\delta_q})}{B(\alpha_q+1,\beta_q+1)+\gamma_q \, B(\alpha_q+\delta_q+1,\beta_q+1) } \,\,\,\,\,\,\,\,\,\,\,\,\,\,\,\,\,\,\,\,\,\,\,\,\,\,\,\, q=u, d
\end{eqnarray}
where $B(\alpha,\beta)$ is the standard beta function,
while for strange quarks {\em plus} anti-quarks we use
\begin{equation}
 x(\Delta s + \Delta \overline{s}) = 2 N_{s}
\frac{x^{\alpha_{{s}}}(1-x)^{\beta_{{s}}}}
{B(\alpha_{{s}}+1,\beta_{{s}}+1)},
\end{equation}
with a similar parametric form for gluons
\begin{equation}
 x\Delta g =N_{g}
\frac{x^{\alpha_{{g}}}(1-x)^{\beta_{{g}}}}
{B(\alpha_{{g}}+1,\beta_{{g}}+1)}.
\end{equation}

The first moments of the quark densities $\delta q$ ($N_q$)
\begin{equation}
\delta q=\int_0^1 dx\, \Delta q
\end{equation}   
 are often related
to the hyperon beta decay constants $F$ and $D$ through the SU(3) symmetry 
relations
\begin{eqnarray} 
\delta u +\delta \overline{u}- \delta d -\delta
\overline{d}  \equiv  N_{u} - N_{d} 
= F+D = 1.2573
\end{eqnarray} 
\begin{eqnarray} 
\delta u +\delta
\overline{u}+ \delta d +\delta \overline{d}- 2(\delta s +\delta
\overline{s})  \equiv  N_{u} + N_{d}  - 4 N_{s} 
 = 3F-D  =  0.579.
\end{eqnarray}
Under such an assumption, the previous equations would strongly constrain the normalization of the
 quark densities. However, as we are not interested in forcing flavor symmetry, we leave aside 
that strong assumption and  relax the symmetry relations introducing two parameters, $\epsilon_{Bj}$ 
and $\epsilon_{SU(3)}$ respectively. These parameters account quantitatively for eventual departures 
from  flavor symmetry considerations, including also some uncertainties on the low-$x$ behavior, and 
 higher order corrections,
\begin{equation} 
\label{Bj}
N_{u} - N_{d} = (F+D)(1+\epsilon_{Bj})  \,\,\,\,\,\,\,\,\,\,\,\,\,\,\,\,\,\,\,\,\,\,\,\,\,\,\,\,\,\,\,\,\,\,\,\,
 N_{u} + N_{d}  - 4 N_{{s}} =(3F-D)(1+\epsilon_{SU(3)}),
\end{equation}
and we take them as a measure of the
degree of fulfillment of the Bjorken sum rule \cite{BJ} and the $SU(3)$ symmetry.

Equations (\ref{Bj}) 
allow to write the normalization of the three quark flavors in terms
of $N_{{s}}$, $\epsilon_{Bj}$, and $\epsilon_{SU(3)}$. Notice that no constraints have been imposed on the 
breaking parameters since we expect them to be fixed by data. The remaining parameters are constrained in such 
a way that positivity with respect to MRST02 parton distributions is fulfilled.
These last parameterizations are used in order to compute the denominators of equations (1) and (3). Similar results are obtained with other modern sets of 
PDFs.
Consistently with the choice for the unpolarized parton distributions, in order to compute $\alpha_s$ at LO and NLO we use the values of $\Lambda_{QCD}$ 
obtained Ref.\cite{MRST02} .

As anti-quark densities we take 
\begin{equation}
 x\Delta \overline{q} =N_{\overline{q}}
\frac{x^{\alpha_{\overline{q}}}(1-x)^{\beta_{\overline{q}}}}
{B(\alpha_{\overline{q}}+1,\beta_{\overline{q}}+1)}\,\,\,\,\,\,\,\,\,\,\,\,\,\,\,\,\,\,\,\,\,\,\,\, \overline{q}=\overline{u},\overline{d}
\end{equation}
for $\overline{u}$ and $\overline{d}$ quarks, and we assume $\overline{s}=s$
since the possibility of discrimination in the $s$ sector is beyond the
precision of the data (as in the unpolarized case).

Fragmentation functions are taken from either \cite{kretzer} or \cite{KKP}, 
respectively. We also use the flavor symmetry and flavor separation criteria 
 proposed in \cite{kretzer}, at the respective initial scales $Q^2_{i}$.
\begin{eqnarray}
&& D_u^{\pi^+}(z,Q^2)=D_{\bar d}^{\pi^+}(z,Q^2) =D_{\bar u}^{\pi^-}(z,Q^2) = D_d^{\pi^-}(z,Q^2)\nonumber \\
&&D_{\bar u}^{\pi^+}(z,Q^2_i) = D_d^{\pi^+}(z,Q^2_i)=
 D_s^{\pi^+}(z,Q^2_i)=D_{\bar s}^{\pi^+}(z,Q^2_i)=D_{ s}^{\pi^+}(z,Q^2_i)=
(1-z)\,D_u^{\pi^+}(z,Q^2_i)
\end{eqnarray}
for the fragmentation into pions, which have shown to be in agreement 
with unpolarized SIDIS data \cite{KE}, and
\begin{eqnarray}
&& D_u^{K^+}(z,Q^2)=D_{\bar u}^{K^-}(z,Q^2)=(1-z)\,D_{\bar s}^{K^+}(z,Q^2)  \nonumber \\
&& D_d^{K^+}(z,Q^2_i)=D_{\bar d}^{K^+}(z,Q^2_i)= D_s^{K^+}(z,Q^2_i)=D_{\bar u}^{K^+}(z,Q^2_i)=
(1-z)^2\,D_{\bar s}^{K^+} (z,Q^2_i)
\end{eqnarray}
for kaons.

The data sets analyzed  include only points with $Q^2>1$ GeV$^2$,
listed in Table \ref{tab:table1},  and totaling 137, 139, and 37 points, from 
proton, deuteron, and helium targets respectively, from pDIS plus 60, 87, and 
18, from proton, deuteron, and helium targets respectively from pSIDIS.
Notice that in the case of HERMES data, we have taken those coming from the 
most recent release, which include corrections for instrumental smearing and 
radiative effects

\begin{table}
\caption{\label{tab:table1} Inclusive and semi-inclusive data used in the fit.}
\begin{ruledtabular}
\begin{tabular}{ccccc} 
Collaboration & Target& Final state & \# points & Refs. \\ \hline
EMC          & proton& inclusive   &    10     & \cite{EMC} \\ 
SMC          & proton, deuteron & inclusive &  12, 12  & \cite{SMCi} \\  
E-143        & proton, deuteron & inclusive  &  82, 82  & \cite{E143} \\ 
E-155        & proton, deuteron & inclusive   &    24, 24    & \cite{E155} \\ 
Hermes       & proton,deuteron,helium& inclusive   &    9, 9, 9   & \cite{HERMES} \\
E-142        & helium& inclusive   &    8     & \cite{E142} \\ 
E-154        & helium& inclusive   &    17     & \cite{E143} \\ 
Hall A       & helium & inclusive &   3      & \cite{HALLA}   \\
COMPASS      & deuteron & inclusive &   12  & \cite{COMPASS} \\
 \hline 
SMC          & proton,deuteron& $h^+$, $h^-$  &  24, 24  & \cite{SMC} \\ 
Hermes       & proton, deuteron, helium & $h^+$, $h^-$, $\pi^+$, $\pi^-$, $K^+$, $K^-$, $K^T$   &    36,63,18     
& \cite{HERMES} \\  \hline
\multicolumn{3}{c}{Total} & 478 & \\
\end{tabular}
\end{ruledtabular}
\end{table}

\section{Global Analyses}

As it is well known, there are various alternatives for defining the function
to be minimized in the global fit \cite{Stump:2001gu,Thorne:2002jx}. To begin with, we consider 
the most simple and commonly used in fits to polarized data, namely 
\begin{equation}
\chi^2=\sum_{i=1}^N \frac{(T_i-E_i)^2}{\sigma_i^2}\,,
\end{equation}
where $E_i$ is the measured value of a given observable, $T_i$ is the
corresponding theoretical estimate computed with a given set of parameters
for the pPDFs, and $\sigma_i$ is the error associated with the measurement,
usually taken to be the addition of the reported statistical and systematic 
errors in quadrature. This definition ignores the correlations between data 
points from the same measurements, nevertheless it is widely used since in 
many cases the full correlation matrices are not available. In the cases 
where the correlation matrices between inclusive and semi-inclusive data are 
available, we have taken them into account by means of 
the appropriate definition for $\chi^2$ \cite{DSS}.

In Table \ref{tab:table2}, we summarize the results of the best NLO and LO 
global fits to all the data listed in Table \ref{tab:table1} (478 data points). We present fits obtained using alternatively fragmentation functions from 
reference \cite{kretzer}, labeled as  KRE, and from reference \cite{KKP}, 
labeled as KKP.
\begin{table}[hbt]
\caption{\label{tab:table2} $\chi^2$ values and first moments for distributions at $Q^2=10$ GeV$^2$}
\begin{tabular}{cccccccccccc} \hline \hline
\multicolumn{2}{c}{set}  & $\chi ^2$ &$\chi^2_{DIS}$ &$\chi^2_{SIDIS}$  & $\delta u_v$ & $\delta d_v$  & $\delta \overline{u}$ & 
$\delta \overline{d}$   & $\delta \overline{s}$ & 
$\delta g$ & $\delta \Sigma$\\ \hline
& KRE & 430.91 &206.01 & 224.90   & 0.936 & -0.344 & -0.0487 &-0.0545 & -0.0508 & 0.680 & 0.284\\ 
\raisebox{1.7ex}{NLO}  & KKP & 436.17&205.66&230.51 & 0.700 & -0.255 & 0.0866 &-0.107 & -0.0454  & 0.574 & 0.311\\ \hline
 & KRE  & 457.54&213.48 &244.06   & 0.697 & -0.248 & -0.0136  &-0.0432  & -0.0415 & 0.121 &
 0.252 \\
\raisebox{1.7ex}{LO}  & KKP  & 448.71&219.72  & 228.99 & 0.555  & -0.188  & 0.0497& -0.0608 & -0.0365 &0.187 & 0.271 \\ \hline \hline
\end{tabular}
\end{table}
Since the fit involves 20 parameters, the number of degrees of freedom for 
these fits is 478-20=458. Consequently, the $\chi^2$ values obtained are 
excellent for NLO fits and very good for LO. The better agreement between 
theory and experiment found at NLO, highlights the importance of the 
corresponding QCD corrections, for the present level of accuracy achieved by the 
data.

In NLO fits there seems to be better agreement when using KRE 
fragmentation functions, whereas at LO the situation is the opposite. 
The difference between the total $\chi^2$ values between KRE and KKP NLO fits 
comes mainly from the contributions related to pSIDIS data, while those 
associated to inclusive data are almost the same, as one should expect in a 
fully consistent scenario. At variance with the NLO situation, in LO fits, the 
effect of pSIDIS data in the fit to inclusive data is strongly dependent on 
the set FFs used. This suggest a flaw in either the LO description of pSIDIS, 
the LO FFs used, or most likely in both of these ingredients. Notice also
that KKP LO fit achieves a  $\chi^2$ to pSIDIS data close to those obtained at 
NLO, but the degree of agreement with inclusive data is several units poorer.
KRE LO fit improves the agreement with inclusive data, but with a higher
 $\chi^2$ to pSIDIS data.

Table  \ref{tab:table2} includes also the first moment of each flavor distribution at $Q^2=10$ GeV$^2$, and that for the singlet distribution $\delta \Sigma$, as reference.
Most noticeably, while the KRE NLO fit favors the idea of a SU(3) symmetric 
sea, KKP NLO finds $\overline{u}$ polarized opposite to $\overline{d}$ and to $\overline{s}$. Gluon and strange sea quark polarization are similar in both fits
and the total polarization carried by quarks is found to be around 30\%.    
At LO, KRE fits also prefers sea quarks polarized in the same direction 
although $\overline{u}$ is much less polarized than $\overline{d}$ and
 $\overline{s}$. KKP LO shows $\overline{u}$ polarized opposite to 
$\overline{d}$ and to $\overline{s}$ as in NLO, and the total polarization 
is around 25\%.

Notice that $\Delta u + \Delta \overline{u}$ and $\Delta d + \Delta 
\overline{d}$, could in principle be determined using inclusive data alone,
with no dependence on a FFs. However, since pSIDIS data determine both
  $\Delta q + \Delta \overline{q}$ and $\Delta \overline{q}$, in the combined 
fit $\Delta q + \Delta \overline{q}$ ends with a small dependence on the FFs. 
The full set of parameters for the different sets can be found in Table A1,
in the Appendix. A Fortran subroutine providing the scale dependent pPDFs
can be obtained upon request from the authors.

\setlength{\unitlength}{1.mm}
\begin{figure}[hbt]
\includegraphics[width=19cm]{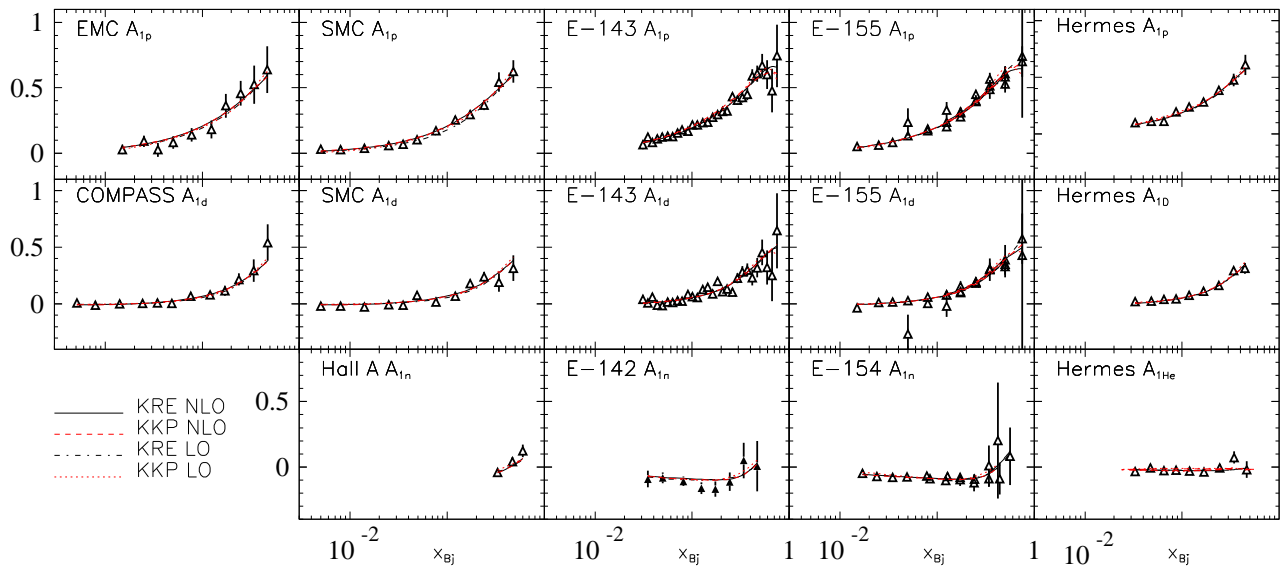}
\caption{Inclusive asymmetries computed with LO and NLO pPDFs against
the corresponding data.}
\label{fig:ainc}
\end{figure}
\setlength{\unitlength}{1.mm}
\begin{figure}[hbt]
\includegraphics[width=19cm]{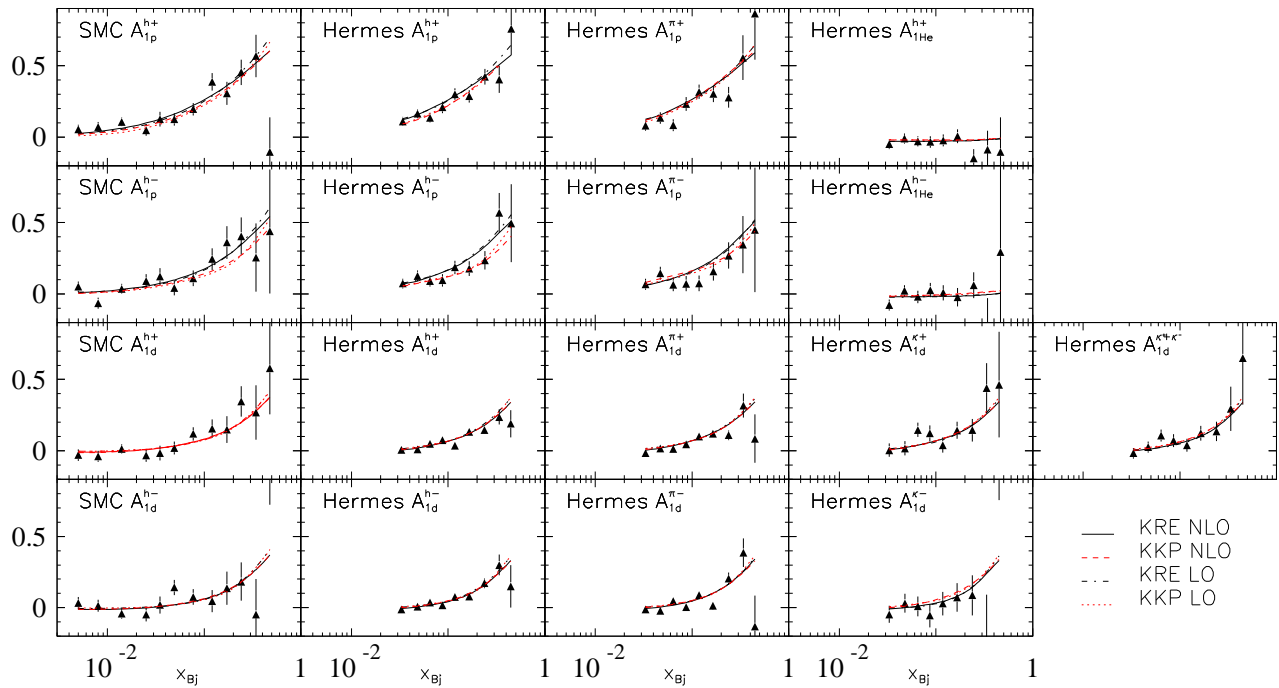}
\caption{Semi-inclusive asymmetries computed with LO and NLO pPDFs against
the corresponding data.}
\label{fig:asinc}
\end{figure}

In Figures \ref{fig:ainc} and \ref{fig:asinc} we show the 
inclusive and semi-inclusive asymmetries computed with the different 
parameterizations both at LO and NLO accuracy, against the corresponding data 
sets. The differences between the various sets can hardly be noticed in the 
comparison to inclusive data in Figure \ref{fig:ainc}, although are more
significant when comparing to pSIDIS data, specially in the case of proton 
targets. This is due to the fact that the main difference between the sets 
are the light sea quark densities, which are probed by pSIDIS processes of 
proton targets. The pSIDIS asymmetries for deuterium targets are, of course, 
less sensitive to these differences since they average 
$\overline{u}$ and $\overline{d}$ contributions.

\section{Uncertainties}

A crucial issue to be addressed before going any further in the 
interpretation of the results of the previous section is to estimate
the uncertainties in the extraction of the individual pPDFs by means of 
the global fit, and also the uncertainty that will have any observable 
computed with them. This has been thoroughly studied in the context of
unpolarized parton distributions, where the number and precision of the data 
available is much more significant, and the correct estimate of the 
uncertainties arising from PDF in predictions for observables related 
to  new physics is mandatory \cite{BOTJE,Stump:2001gu,MRST02}.  

The sources of uncertainty in PDFs are often classified into those associated 
with experimental errors on the data, and those associated with rather 
theoretical or phenomenological assumptions in the global fitting procedure, 
including: higher order QCD effects in the analyzed cross sections and their 
scale dependence, the particular choice of the parametric form of the 
distributions at the initial scale, nuclear target corrections, 
hadronization mechanisms and model assumptions such as $s=\overline{s}$.  
Clearly, while the first category is usually under control, the second 
one is particularly difficult to determine.

Many strategies have been implemented in order  to assess the uncertainties in 
PDFs and their propagation to observables, specially those associated with experimental errors in the data. These
include the Hessian approach, which assumes that the deviation in $\chi^2$ 
for the global fit is quadratic in the deviations of the parameters specifying 
the input parton distributions, and then propagates linearly these 
uncertainties  from PDFs to observables. Alternatively, the Lagrange 
multiplier method \cite{Stump:2001gu} probes the uncertainty in any observable 
or quantity of interest much more 
directly. This last method relates the range of variation of one
or more physical observables dependent upon PDFs to the variation in the 
$\chi^2$ used to judge the goodness of the fit to data. 
Specifically, it can be implemented minimizing the function
\begin{equation}
\Phi(\lambda_i, a_j)=\chi^2(a_j)+\sum_i \lambda_i\,O_i(a_j)
\end{equation}
with respect to the set of parameters $a_j$ of the starting PDFs,
for fixed values of the Lagrange multipliers $\lambda_i$. Each one of the parameters $\lambda_i$ is related to
an observable $O_i$, which is computed from the set of parameters $a_j$.
For $\lambda_i=0$ we get the best standard global fit, for which $\chi^2(a_j)=
\chi^2_0$ and $O_i(a_j)=O_i^0$. Varying $\lambda_i$, and minimizing
$\Phi(\lambda_i, a_j)$, the fit to data deteriorates increasing $\chi^2(a_j)$
from its minimum while $O_i(a_j)$ variates due to the different set of 
parameters $a_j$ found. Performing a series of global fits for different
values of $\lambda_i$, we get a profile of  $\chi^2(a_j)$ for a range of values
of the quantities $O_i$. In other words, this tell us how much the fit to 
data deteriorates if we force the PDFs to yield a prediction for an 
observable different to the one obtained with the best fit $O_i^0$.

\begin{figure}[hbt]
\includegraphics[width=18cm]{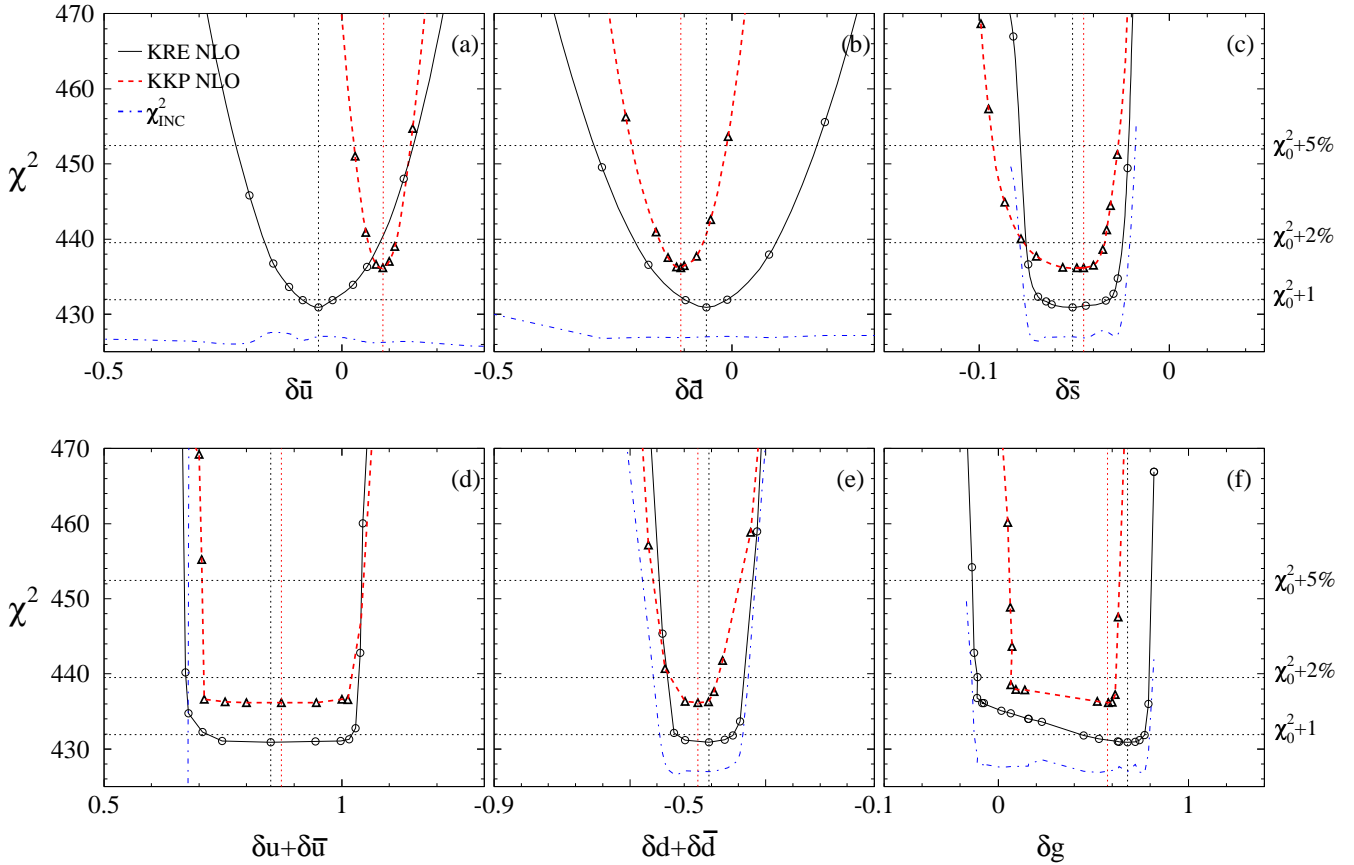}

\caption{$\chi^2$ profiles for NLO fits obtained using Lagrange multipliers
at $Q^2=10$ GeV$^2$.}
\label{fig:krenlo}
\end{figure}

 In order to illustrate the method and apply it to fits of polarized data, in 
Figure \ref{fig:krenlo}  we show the outcome of varying the $\chi^2$ of the 
NLO fits to data against the first moment of the respective polarized parton 
densities $\delta q$ at $Q^2=10$ GeV$^2$, one at a time. This is, to minimize
\begin{equation}
\Phi(\lambda_q, a_j)=\chi^2(a_j)+\lambda_q\, \delta q(a_j) \,\,\,\,\,\,\,\,\,\,\,\, q=u,\overline{u},d,\overline{d},s,g.
\end{equation}
Of course, beyond LO,
these first moments are not, strictly speaking, physical observables, however
are perfectly well defined quantities once the factorization prescription is 
fixed. The solid lines in Figure \ref{fig:krenlo}  correspond to the KRE NLO fit (using as input for fragmentation
functions those of reference \cite{kretzer}) while the dashed lines 
correspond KKP NLO (using \cite{KKP}).

\setlength{\unitlength}{1.mm}

In and {\em ideal} situation, i.e. to have reliable estimates of every source 
of uncertainty, of correlated experimental and theoretical errors, and a 
quadratic dependence of $\chi^2$ in the parameters of the fit, the profile 
of $\chi^2$ would be just a parabola and 
the 1$\sigma$ uncertainty in any observable would correspond to $\Delta 
\chi^2=1$. In order to account for unexpected sources of uncertainty, in 
modern unpolarized global analysis it is customary to consider instead of 
$\Delta \chi^2=1$ between a 2\% and a 5\% variation in $\chi^2$ as 
conservative 
estimates of the range of uncertainty.

As expected in the {\em ideal} framework, the dependence of $\chi^2$ on the 
first 
moments of $\overline{u}$ and $\overline{d}$ resemble a parabola (Figures 3a and 3b). 
The KKP 
curves are shifted upward almost six units relative to those from KRE, due 
to the difference in $\chi^2$ of their respective best fits. 
Although this means that the overall goodness of KKP fit is poorer than KRE, 
$\delta \overline{d}$ and $\delta \overline{u}$ seem to be more tightly 
constrained. The estimates for $\delta \overline{d}$  computed with the 
respective best fits are close and within the $\Delta \chi^2=1$ range, 
suggesting something close to the {\em ideal} situation. However for 
$\delta \overline{u}$, 
they only overlap allowing a variation in $\Delta\chi^2$ of the order of a 
2\%. This is a very 
good example of how the $\Delta \chi^2=1$ does not seem to apply due to an 
unaccounted source of uncertainty: the differences between the available sets 
of fragmentation functions. 
\setlength{\unitlength}{1.mm}
\begin{figure}[hbt]
\includegraphics[width=18cm]{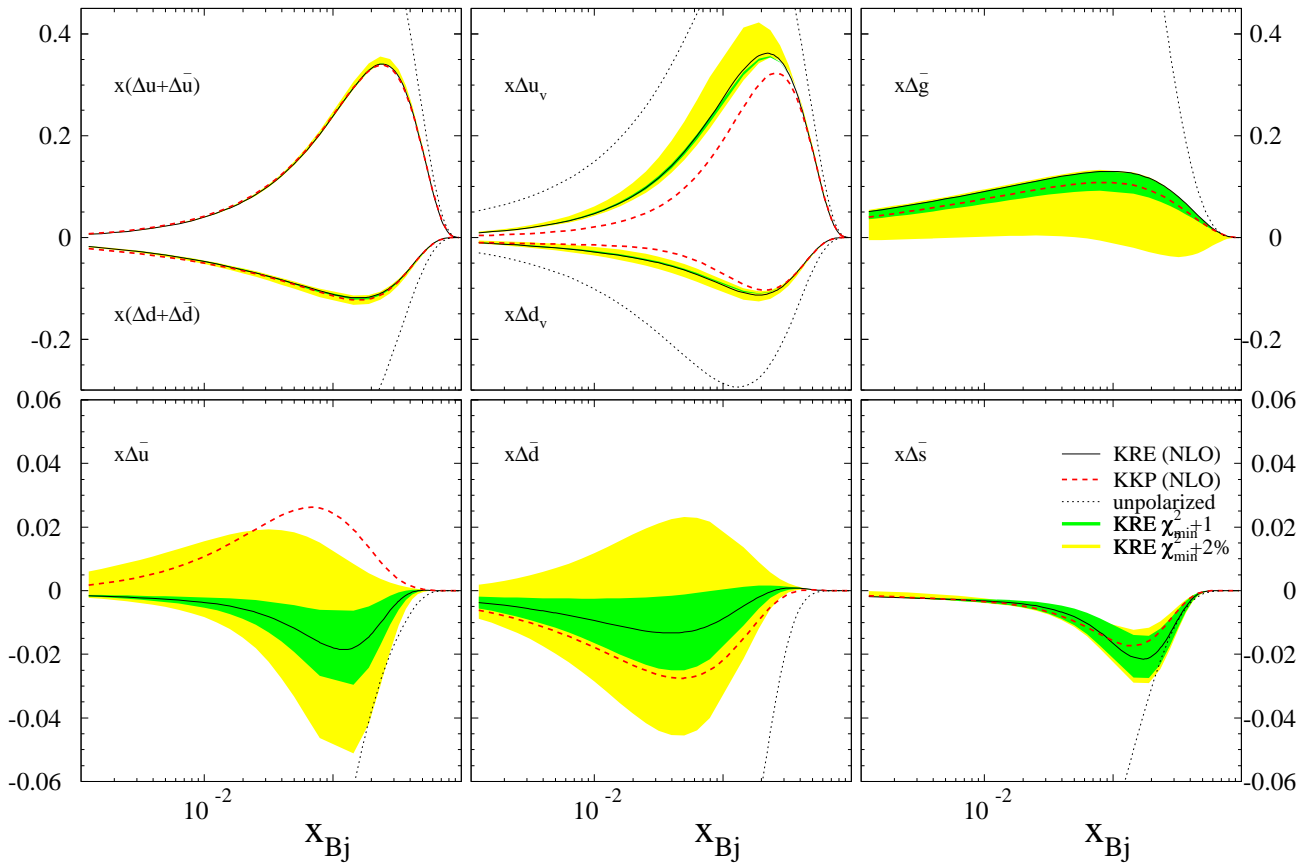}
\caption{Parton densities at $Q^2=10$ GeV$^2$, and the uncertainty bands corresponding to $\Delta \chi^2=1$ and $\Delta \chi^2=2\%$  }
\label{fig:disnlo}
\end{figure}

An interesting thing to notice is that almost all the variation in $\chi^2$
comes from the comparison to pSIDIS data. The partial  $\chi^2$ value computed only
with inclusive data, $\chi^2_{pDIS}$, is almost flat reflecting the fact the
pDIS data are not sensitive to $\overline{u}$ and $\overline{d}$ distributions.
In Figure \ref{fig:krenlo},  we plot $\chi^2_{pDIS}$ with an offset of 206 units as a dashed-dotted line.

The situation however changes dramatically when considering 
$\delta \overline{s}$ or $\delta g$ as shown in Figures 3c and 3f, respectively. 
In the case of the variation 
with respect to $\delta \overline{s}$, the profile of $\chi^2$ is not at all 
quadratic, and the distribution is much more tightly constrained (notice 
that the scale used for $\delta \overline{s}$ is almost four times smaller 
than the one used for light sea 
quarks moments). The  $\chi^2_{pDIS}$ 
corresponding to inclusive data is more or less indifferent within an interval 
around the best fit value and increases rapidly on the boundaries. This steep 
increase is related to a positivity constraints on $\Delta s$ and $\Delta g$. 
pSIDIS data have a similar effect but also helps to define a minimum within the 
interval. The preferred values for $\delta s$ obtained from both NLO fits are 
very close, and in the case of KRE fits, it is also very close to those 
obtained for $\delta \overline{u}$ and $\delta \overline{d}$ suggesting SU(3) 
symmetry. 

For the total $u$ and $d$  polarization $\delta u + \delta \overline{u}$ and
$\delta d + \delta \overline{d}$, the absolute uncertainties are large as shown 
in Figures 3d and 3e respectively,  
however their values relative to the best fit results are significantly smaller
than for sea quarks.  
For $\delta g$, again $\chi^2_{pDIS}$ just defines an interval, as 
it has been pointed out in previous analyses \cite{DSS,DS}, bounded by its own
positivity constraint in the upper end, and by those of the sea quarks 
densities, which grow in the lower end.  Between both ends, there is either 
some redundancy of parameters or a need of more sensitive data. Again, 
semi-inclusive data help to define a minimum, and it is very close for both 
fits. Within the $\Delta \chi^2=2\%$ uncertainty, the gluon polarization
estimate is in agreement with a recent analysis of high $p_T$ hadron production
\cite{Ellis:2005cy}

Clearly, in the case of  $\delta \overline{s}$ or $\delta g$, the 
simple minded use of the Hessian approach does not apply. In the case of
the light sea, the sensitivity to a smaller and internally consistent
data set, presumably put us closer to the ideal situation.

In order to see the effect of the variation in $\chi^2$ on the parton 
distributions themselves, in Figure  \ref{fig:disnlo}, we show KRE best fit
densities together with the uncertainty bands corresponding to 
$\Delta \chi^2=1$ (darker band) and $\Delta \chi^2=2\%$ (light shaded band). 
As expected, the relative uncertainties in the total quark densities and 
those strange quarks are rather small. For gluon densities the 
$\Delta \chi^2=1$ band is also small, but the most conservative 
$\Delta \chi^2=2\%$ estimate is much more significative. For light sea quarks 
the  $\Delta \chi^2=1$ bands are moderate but the $\Delta \chi^2=2\%$ are much more larger.

\setlength{\unitlength}{1.mm}
\begin{figure}[hbt]
\includegraphics[width=18cm]{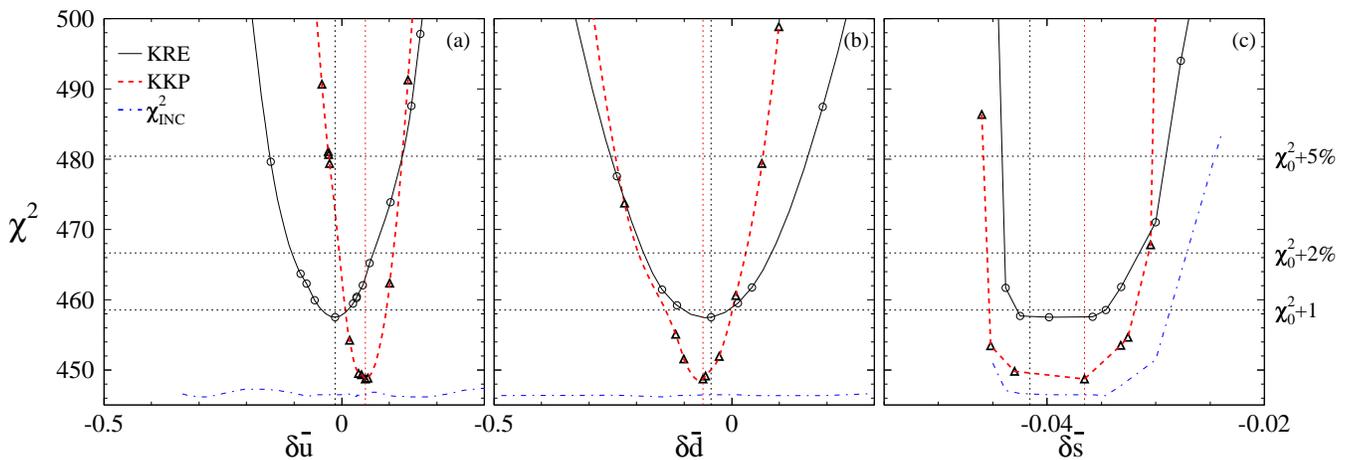}
\caption{$\chi^2$ profiles for LO fits obtained using Lagrange multipliers
at $Q^2=10$ GeV$^2$.}
\label{fig:krelo}
\end{figure}

The profiles for $\chi^2$ obtained with LO fits show similar features to 
those obtained at NLO. In Figure  \ref{fig:krelo} we show the LO profiles 
for sea quark densities. Notice that before reaching any conclusion about 
the uncertainties obtained in LO and their comparison with those found in 
NLO a few remarks are in order. In the first place, the rather large 
difference between the LO and NLO $\chi^2$ values in fits to the same sets of 
data implies that the uncertainties due to use of one or the other 
approximation, and their related ingredients, are not properly accounted for. 
Any mean to include these uncertainties would certainly reduce the 
constraining power of the LO fit an thus increase the uncertainties on 
the pPDFs. Unfortunately 
there is no evident way to do this and also to relate the criteria used
to go from profiles to uncertainties in NLO and LO. The second point is 
related to the fact that beyond the LO PDFs are factorization scheme 
dependent and thus the relation between the LO and NLO PDFs and their 
uncertainties is even less direct.

The Lagrange multiplier method allows also to analyze the interplay and 
consistency between inclusive and semi-inclusive data \cite{NS}  
we mentioned in the previous section. The idea is to apply a Lagrange 
multiplier to the pSIDIS contribution to the function to be minimized, 
\begin{equation}
\label{chisemi}
\Phi(\lambda, a_j)=\chi^2_{pDIS}(a_j)+\lambda\,\chi^2_{pSIDIS}(a_j) 
\end{equation}
and perform global fits for different values of $\lambda$. For $\lambda$ values
lower than unity, the weight of pSIDIS data in the fit is artificially reduced
and the fit becomes increasingly dominated by pDIS data as $\lambda \rightarrow 0$. The partial contribution of pDIS data to $\chi^2$, $\chi^2_{pDIS}$ 
decreases up to a saturation point given by the best fit to pDIS data only, 
while that of pSIDIS data, $\chi^2_{pSIDIS}$, increases. Using $\lambda$ values
larger than unity, the fit becomes dominated by pSIDIS data, until it 
saturates at the best fit to pSIDIS data alone.  
 
\setlength{\unitlength}{1.mm}
\begin{figure}[hbt]
\includegraphics[width=8cm]{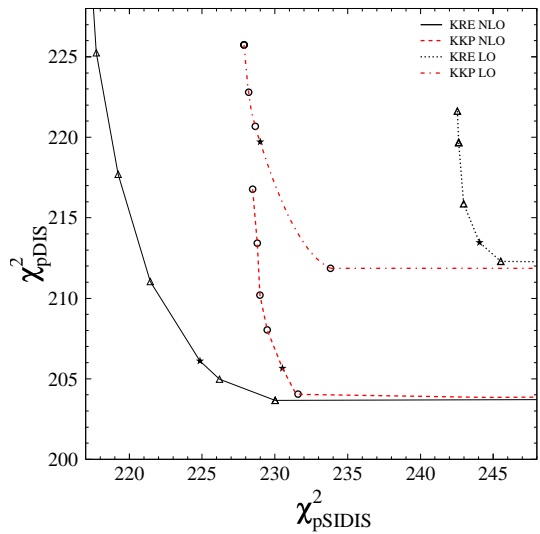}
\caption{$\chi^2_{pDIS}$ vs. $\chi^2_{pSIDIS}$ for KRE and KKP NLO and LO fits}
\label{fig:seminlo}
\end{figure}

In Figure \ref{fig:seminlo} we show the results for $\chi^2_{pDIS}$ vs. $\chi^2_{pSIDIS}$ in fits with a wide range of values of  $\lambda$ in Eq. 
\ref{chisemi} for KRE NLO (solid line), KKP NLO (dashed line), KRE LO 
(dotted line) and KKP LO fits (dotted-dashed line), respectively. The 
standard best fit in each curve ($\lambda=1$) is denoted by a star. 

In both NLO curves, the $\chi^2_{pDIS}$ values obtained in the best 
standard fits, are very close to the common saturation value, 
corresponding to a fit without pSIDIS data. This suggests a high degree of
compatibility between both data sets in a NLO framework, since
the inclusion of pSIDIS data does not worsens significantly the agreement 
with pDIS data. The increase in the relative weight of pSIDIS data 
reduces $\chi^2_{pSIDIS}$ only in a couple of units in the case KKP fits, 
what indicates that the fit is strongly constrained and that the 
best fit result for $\chi^2_{pSIDIS}$ is close to its saturation value. 
However, $\chi^2_{pSIDIS}$ in the best KRE NLO is several units smaller 
although with a $\chi^2_{pDIS}$ similar to that of KKP. The saturation 
value obtained in this case is also much smaller. These last features 
mean that KRE FFs allow a much better and flexible fit to pSIDIS data.

For LO the situation is quite different: the common saturation value for
$\chi^2_{pDIS}$ is several units larger that the found at NLO, what means
that even neglecting pSIDIS data the fits obtained at LO are poorer than 
those of NLO accuracy. The best KRE LO fit value for $\chi^2_{pDIS}$ is 
close to the saturation value, but at the expense of a rather large 
$\chi^2_{pSIDIS}$ value. In the case of KKP LO fit, $\chi^2_{pSIDIS}$ 
improves significantly but the departure of $\chi^2_{pDIS}$ from its 
saturation value is very large.

\section{Future prospects}

One of the measurements most eagerly awaited by the spin physics community
is that of single inclusive large $p_T$ pion production in longitudinally 
polarized proton-proton collisions, which is right now being run at BNL RHIC
\cite{Adler:2004ps}. 
The spin dependent asymmetry associated to this kind of process, 
$A^{\pi^0}_{LL}$ is defined, as usual, in terms of the ratio between the 
polarized and the unpolarized cross sections,
\begin{equation}
A^{\pi^0}_{LL}=\frac{d\Delta \sigma^{p\,p\rightarrow\pi^0\,X}}
{d \sigma^{p\,p\rightarrow\pi^0\,X}}
\end{equation}
which is strongly dependent on the gluon polarization. Indeed, in this 
observable polarized gluons show up in the cross section in the dominant 
terms. The NLO QCD corrections to this observable have been computed recently 
\cite{deFlorian:2002az,Jager:2002xm} finding both large QCD corrections, 
 which make unavoidable the use of the NLO approach, and  also a very 
significant dependence on $\Delta g$, features that in principle would help
to constrain the amount of gluon polarization in much more stringent way
than in DIS experiments.

The data obtained up to now by the PHENIX Collaboration suggest a very small
asymmetry, consistent with pPDFs sets with a moderate gluon polarization.
In the following we apply the Lagrange multiplier method in order to explore
the range of variation of the estimates for this asymmetry associated to the 
uncertainty in the present extraction of pPDFs.

\setlength{\unitlength}{1.mm}
\begin{figure}[hbt]
\includegraphics[width=8cm]{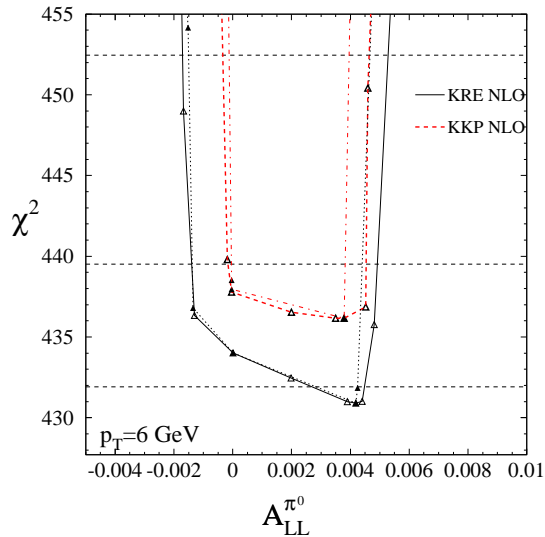}
\caption{Profile of the global $\chi^2$ to data against 
$A^{\pi^0}_{LL}$ at $p_T=6$ GeV}
\label{fig:asy}
\end{figure}

In Figure \ref{fig:asy} we show the range of variation of 
$A^{\pi^0}_{LL}$ at an intermediate value for $p_T=6$ GeV obtained with 
different sets of pPDFs 
against the variation of the $\chi^2$ to pDIS and pSIDIS data for these 
distributions. The profile of $\chi^2$ defines a well defined range of values 
for $A^{\pi^0}_{LL}$ allowed by present pPDFs. The solid line represents 
the profile of $\chi^2$ obtained using KRE FFs, both
in the global fit to data and in the computation of the asymmetry. The dashed
line represents the same but for KKP FFs. The minima corresponding to both profiles
are very close suggesting a cancellation of the associated uncertainty. 
Notice that both the extraction of the pPDFs, and also the estimate of 
$A^{\pi^0}_{LL}$ with a given set, rely on a set of FFs. The double 
dependence of the observable on the set of FFs used may had, in principle, 
even potentiated the disagreement.

Notice that when the fitting procedure explores alternative
sets of pPDFs in order to  minimize or maximize the value of 
$A^{\pi^0}_{LL}$, not only the gluon distributions varies, but all the 
distributions. Since the observable is strongly dependent on the gluon
polarization, there is a very tight correlation between $A^{\pi^0}_{LL}$
and $\delta g$, reflected by the fact that almost the same sets that 
maximize/minimize the asymmetry, maximize/minimize $\delta g$, however the
role of sea quark polarization is non negligible.

In Figure \ref{fig:asy} we have included also the profiles obtained using 
$\delta g$ instead of $A^{\pi^0}_{LL}$ in minimization as a dotted line in the 
case of KRE and a dashed dotted line for KKP. Clearly, the sea quark 
polarization can conspire in order to yield a larger/smaller asymmetry than 
the one obtained with maximum/minimum gluon fit at a given $\chi^2$, effect 
which in much more apparent at larger values of $A^{\pi^0}_{LL}$. This feature
will have to be taken into account in the future for a very precise 
measurement of the gluon polarization.
\setlength{\unitlength}{1.mm}
\begin{figure}[hbt]
\includegraphics[width=8cm]{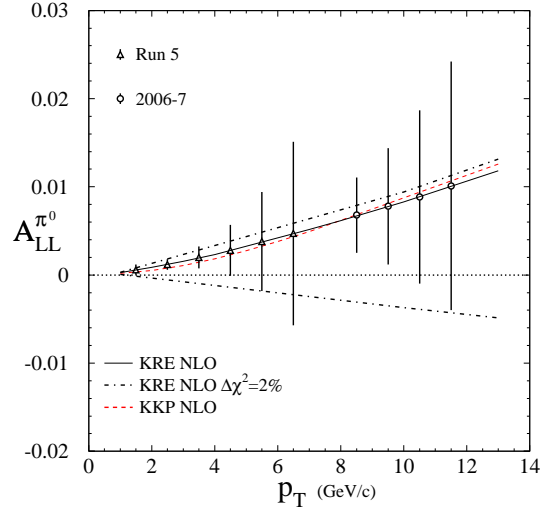}
\caption{
$A^{\pi^0}_{LL}$ vs. $p_T$ computed with different pPDF sets and
the estimated errors expected by PHENIX at RHIC.}
\label{fig:asy3}
\end{figure}
\setlength{\unitlength}{1.mm}
\begin{figure}[hbt]
\includegraphics[width=14cm]{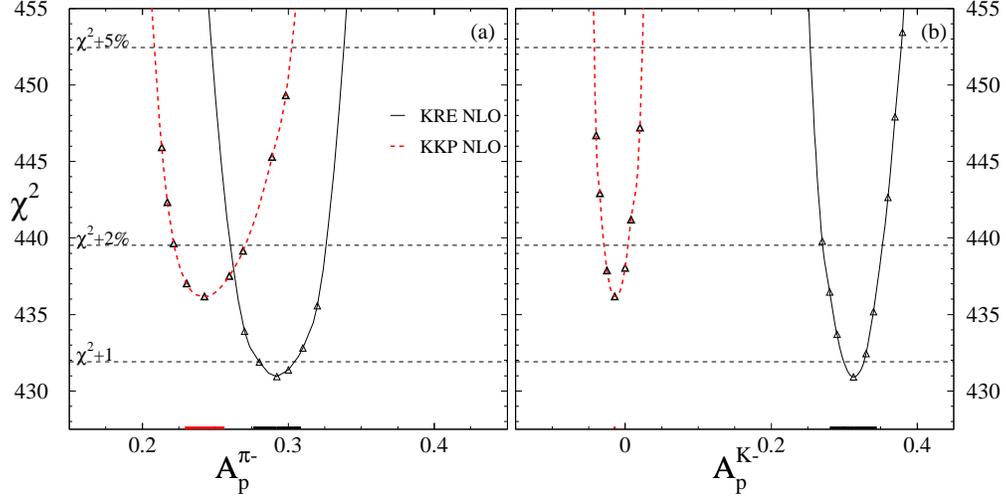}
\caption{Profile of the global $\chi^2$ to data against 
$A^{\pi^-}_p$ and $A^{K^-}_p$ proposed to be measured by E04-113 at JLAB.}
\label{fig:apik}
\end{figure}

Unfortunately, the data collected so far in the first two runs of the PHENIX
detector at RHIC have estimated errors much larger than the uncertainties in 
the values $A^{\pi^0}_{LL}$ coming from the fits, however this situation 
is going to change dramatically in the near future. In Figure \ref{fig:asy3} 
we plot  $A^{\pi^0}_{LL}$ as a function of $p_T$ using the 
best NLO fits coming from KRE and KKP FFs (solid lines and dashes, 
respectively) and also with KRE variants designed to enhance/reduce 
$A^{\pi^0}_{LL}$ with $\Delta \chi^2= 2\%$ (dotted-dashed lines). 
We have also included the expected uncertainties for the next two runs centered
at KRE best fit estimate \cite{Fukao:2005id}. 
Clearly, the future measurement of the asymmetry 
would certainly be able to constrain even further the gluon and therefore 
reduce the uncertainties in pPDFs.

Another source of complementary information to further constrain the 
extraction of pPDFs and also FFs is the experimental program of the 
E04-113 experiment at JLAB, which propose to measure pion and kaon pSIDIS 
asymmetries for proton and deuteron targets \cite{xiaodong}.
 In Figure \ref{fig:apik}a we show the profile of $\chi^2$ of the global fits
using KRE (solid line) and KKP (dashes) FFs against the $\pi^-$ pSIDIS 
asymmetry on proton targets at one of the kinematic configurations of E04-113 
($x_Bj=0.203$, $Q^2=2.3$ GeV$^2$  $<z>=0.5$). The expected uncertainty for the 
measurement of this asymmetry shown in the bottom of the figure, 
is significantly smaller than those of the previous measurements, and also 
smaller than the present uncertainty coming from the fit. 
In Figure \ref{fig:apik}b we show the same as in the previous figure but
but for negative kaons. This asymmetry has not been measured yet and 
the difference between the predictions coming form  KRE (solid line) and KKP 
(dashes) sets is much larger that the expected uncertainties. 
The measurement of this last observable together with the combined effect of 
data for different targets (proton and deuteron) and final state hadrons 
(positive and negative pions and kaons) will certainly constrain the fit even 
further, specifically the sea quark densities, and at the same time provide 
a more stringent test on the quark flavor separation for the FFs used in the 
analysis.

\section{Conclusions}

The availability of an important set of new data on polarized processes
together with the appropriate theoretical tools required to interpret them,
obtained and developed in the last few years, have led the extraction of pPDFs
in the proton to mature and to become an important source of information which 
will still keep growing in the near future. 

In this paper, we have assessed the feasibility of obtaining pPDFs, with 
special emphasis on the sea quark densities, from a combined NLO QCD 
analysis of pDIS and pSIDIS data. We have estimated the uncertainties
associated to the extraction of each parton density, finding a well 
constrained scenario. pSIDIS data is not only consistent with pDIS, but
improves the constraining power of the fit for all the distributions, 
being crucial for the light sea quarks.  

Different choices for the FFs used in the analysis of pSIDIS data lead
to differences in the  light sea quark pPDFs, although these differences are 
within a conservative uncertainty estimate for them. While KRE FFs favor
a SU(3) flavor symmetric sea, KKP FFs suggest an SU(3) broken one, however
the former led to fits of much better quality.

The first moment of the gluon distribution  at a $10$ GeV$^2$ is found to be 
close to 0.6, constrained to be smaller than 0.8 and larger than -0.05 within 
a conservative $\Delta \chi^2=2\%$ range. The upper constraint comming from
the requirement of positivity of the gluon density and the lower one
due to the QCD evolution which forces sea quark densities to saturate
their own positivity constraints.

The overall picture found for the quark densities at a $10$ GeV$^2$ is one in 
which, within uncertainties, up quarks are almost $100\%$ polarized parallel 
to the proton, down quarks anti-parallel in a similar proportion, and sea 
quarks have a small flavor symmetric negative polarization. 

Two programmed experiments, the one based on the PHENIX detector already 
running at RHIC, and the E04-113 experiment at JLab  will be able to 
reduce dramatically the uncertainty in both the gluon and the light sea 
quark densities respectively, the latter providing also an even more
stringent test on fragmentation functions.

\section{Acknowledgements}

 We warmly acknowledge W. Vogelsang, S. Kretzer, 
C. A. Garc\'{\i}a Canal and Xiaodong Jiang for comments and suggestions, 
Y. Fukao for providing PHENIX error estimates.

\pagebreak

\noindent{\Large \bf Appendix: Parameters of the fits}\\

We present here the parameters of the best fits found in our analyses, both at 
LO and NLO.

{\normalsize
\begin{center}
\begin{tabular}{|c|c|c|c|c|} \hline \hline
Parameter  & KRE NLO   & KKP NLO  & KRE LO  & KKP LO    \\ \hline
\hline
 $\epsilon_{BJ}$ 
           & -0.0067  & 0.0280  & -0.1994 & -0.2413  \\ \hline
$\epsilon_{SU(3)}$ 
           & -0.0122  & -0.0795 & -0.15   & -0.1  \\ \hline

$\alpha_u$ & 1.2024   & 1.0908  & 0.8376  & 1.1325 \\ \hline
$\beta_u$  & 3.4517   & 3.2909  & 3.5832  & 4.3992 \\ \hline
$\gamma_u$ & 7.4178   & 9.2128  & 10.270  & 13.287  \\ \hline
$\delta_u$ & 1.0722   & 0.9956  & 1.4483  & 2.2157   \\ \hline

$\alpha_d$ & 0.6717   & 0.4913  & 0.6431  & 0.8770 \\ \hline
$\beta_d$& 4.7090     & 4.3816  & 4.9575  & 5.4356 \\ \hline
$\gamma_d$ & 14.999   & 19.999  & 14.999  & 14.999  \\ \hline
$\delta_d$ & 1.5666   & 1.3891  & 1.8938  & 2.4361  \\ \hline

$N_s$    & -0.0441    &-0.0398  &-0.0367  &-0.0382 \\ \hline
$\alpha_s$ & 3.500    & 2.5000  & 2.1045  & 3.4466   \\ \hline
$\beta_s=\beta_{\overline{u}}=\beta_{\overline{d}}$ 
& 11.741    & 9.7131  & 9.8861  & 14.997 \\ \hline
$N_{\bar{u}}$&-0.0444 &0.0799   &-0.0122  & 0.0453 \\ \hline
$\alpha_{\bar{u}}$&2.500&1.1332 & 2.5828  & 1.1892  \\ \hline

$N_{\bar{d}}$& -0.0454&-0.0971  &-0.0408  &- 0.0557 \\ \hline
$\alpha_{\bar{d}}$&0.9970&0.8979& 0.73443 & 1.0639  \\ \hline

$N_g$      & 0.1273   & 0.0781  & 0.1965  & 0.082 \\ \hline
$\alpha_g$ & 2.398    & 2.2901  & 0.0390  & 1.320  \\ \hline
$\beta_g$ & 2.1398    & 1.5698  & 2.4051  & 4.485   \\ \hline \hline
$\chi^2_{Total}$ &  430.91 & 436.17 & 457.52& 448.71\\ \hline
$\chi^2_{DIS}$ &  206.01 & 205.66 & 213.48 & 228.99\\ \hline
$\chi^2_{SIDIS}$ &  224.90 & 230.51 & 219.72 & 228.99\\ \hline
\hline \end{tabular}
\end{center}}
\begin{center}
{\footnotesize {\bf Table A1}: Coefficients for NLO and LO best fits}
\end{center}

\end{document}